\documentclass[prl,showpacs,twocolumn,preprintnumbers,amsmath,amssymb,superscriptaddress]{revtex4}
\usepackage{graphicx}
\usepackage{dcolumn}
\usepackage{bm}

\begin{document}
\title{Double coherence resonance in neuron models driven by discrete correlated noise}

\author{Thomas Kreuz}
\email{thomas.kreuz@fi.isc.cnr.it} \affiliation{Istituto dei Sistemi Complessi  - CNR, Sesto Fiorentino, Italy}
\author{Stefano Luccioli}
\affiliation{Istituto dei Sistemi Complessi  - CNR, Sesto Fiorentino, Italy} \affiliation{Istituto Nazionale di
Fisica Nucleare, Unit\'a di Firenze, Sesto Fiorentino, Italy}
\author{Alessandro Torcini}
\affiliation{Istituto dei Sistemi Complessi  - CNR, Sesto Fiorentino, Italy} \affiliation{Istituto Nazionale di
Fisica Nucleare, Unit\'a di Firenze, Sesto Fiorentino, Italy}

\date{\today}

\begin{abstract}
We study the influence of correlations among discrete stochastic excitatory or inhibitory inputs on the response
of the FitzHugh-Nagumo neuron model. For any level of correlation the emitted signal exhibits at some finite
noise intensity a maximal degree of regularity, i.e., a coherence resonance. Furthermore, for either inhibitory
or excitatory correlated stimuli a {\it Double Coherence Resonance} (DCR) is observable. DCR refers to a
(absolute) maximum coherence in the output occurring for an optimal combination of noise variance and
correlation. All these effects can be explained by taking advantage of the discrete nature of the correlated
inputs.
\end{abstract}



\newcommand{\abb}{\small\sf}

\maketitle

Excitable systems driven by fluctuations exhibit a large variety of phenomena where noise plays a constructive
role. Among the most studied are \emph{stochastic resonance}, the enhanced detectability of weak periodic
stimuli for an intermediate noise level~\cite{Gammaitoni98}, and \emph{coherence resonance} (CR), the
regularization of the system response at an optimal noise intensity without any external drive~\cite{Gang93,
Pikovsky97}. CR has been observed in many theoretical and experimental setups such as electronic devices,
semiconductor lasers, and climate models (for a recent review cf.~\cite{Lindner04}). In neuroscience evidence of
CR has been reported for the cat's spino-cortical activity~\cite{Manjarrez02} as well as for a variety of single
neuronal models. A prominent example is the two-dimensional FitzHugh-Nagumo (FHN) model~\cite{FitzHugh61} that
incorporates essential neuronal properties such as threshold dynamics (activation) and refractoriness.

A second type of CR with respect to the level of correlation has been observed in research areas as diverse as
laser dynamics~\cite{Buldu01}, digital circuits~\cite{Brugioni05}, chemical reactions \cite{Beato05} and
neuronal models \cite{Casado97}. In computational neuroscience continuous noise is replaced by a series of
discrete random kicks representing the post-synaptic potentials (PSPs) released by excitatory and inhibitory
synapses \cite{Shadlen98, Salinas00, Feng01, Moreno02, Rudolph01}. Correlations are then introduced via the
method of shared input (cf. \cite{Shadlen98, Salinas00, Rudolph01}) such that the correlation equals the
probability that different synapses deliver PSPs at the same time. Whereas the classical CR with respect to the
noise intensity can be explained by the different dependencies of slow activation and fast excitation on the
noise variance~\cite{Rappel94, Pikovsky97, Pradines99} so far the origin of the second type of CR has remained
unclear.

In this Letter we address this issue by investigating a FHN model driven by a large number of stochastic
synaptic inputs with correlations only among trains of either excitatory or inhibitory PSPs. In both cases the
system exhibits the classical CR at any level of correlation, whereas for almost all noise strengths the second
type of CR can be observed. The discrete nature of the inputs allows to vary independently correlation strength
and noise intensity and thus to disclose the different mechanisms responsible not only for the CRs but also for
the {\it Double Coherence Resonance} (DCR, i.e., the existence of an optimal combination of noise and
correlation strength for which the system responds with maximal coherence) discovered for both excitatory and
inhibitory correlation.
%
%

The FHN model can be written as
\begin{equation} \label{Eq:FHN-Model-Pik}
{\dot V} = \phi (V - \frac{V^{3}}{3} - W)  \quad, \quad {\dot W} = V + a - I(t) \enskip,
\end{equation}
with a voltage variable $V$, a recovery variable $W$ and $I(t)$ the external synaptic input. We set time scale
separation $\phi=100$ and bifurcation parameter $a = a_0 \equiv 1.05$ so that the dynamics has only one
attractor (a stable focus) but is close to the supercritical Hopf bifurcation at $a=1$.

In this study we consider a balanced FHN neuron in the {\it high-input regime} \cite{Shadlen98} with the input
modelled as the superposition of an equal number $N \sim 100 - 10000$ of Poissonian trains of excitatory and
inhibitory PSPs (EPSPs and IPSPs, respectively) with the same rate $\nu_0$
\begin{equation} \label{Eq:Spike-Train-Current}
I(t) = \Delta W_0 \Big[\sum_{i=1}^{N}\sum_{j} \delta(t-t_{i}^{j}) -\sum_{k=1}^{N}\sum_{l}
\delta(t-t_{k}^{l})\Big]
\end{equation}
\noindent where $t_{i}^{j}$ ($t_{k}^{l}$) are the times of the instantaneous excitatory (inhibitory) kicks of
amplitude $\Delta W_0$. The neuron fires upon a single excitatory kick with amplitude higher than $(\Delta W)_c
\simeq 0.0138$ (for $a=a_0$). However, here we consider kicks of much smaller amplitude $\Delta W_0=0.0014$ with
rates $\nu_0 = 0.3$, $0.6$, and $1.2$ comparable with the firing rate $\nu_{f}=\nu_{f}(a)$ just beyond the
bifurcation~\footnote{We use a fourth order Runge-Kutta integration scheme with time step $\delta t= 10^{-4}$. A
spike is identified whenever $W(t)$ overcomes a fixed detection threshold $\Theta=0.4$.}.

We examine the influence of input correlation $\rho_x$ on the neuronal response restricting ourselves to
correlation among either excitatory ($x=e$) or inhibitory ($x=i$) inputs only. Correlations are expressed in
terms of the Pearson correlation coefficient $\rho_x$ equal to the average fraction of shared kicks delivered by
each pair of synapses or, analogously, to the average fraction of synapses delivering kicks at the same time
\cite{Salinas00}. Since in our model the inputs of different synapses are not distinguishable their effect can
be reproduced by two over-all kick trains corresponding to correlated and uncorrelated PSPs, respectively. While
the superposition of the uncorrelated kick trains can be modelled as a single Poissonian sequence of PSPs of
constant amplitude $\Delta W_0$ and rates $\nu_u =N \nu_0 \sim 5-1200$ quite high with respect to the natural
firing frequency, correlated PSPs are generated using a refined method of shared inputs (cf. \cite{Shadlen98,
Salinas00, Rudolph01}). The superposition of the correlated kick trains (with correlation $\rho_x$) can be
represented as a unique over-all Poissonian train of kicks of variable amplitude $\Delta W = n \times \Delta
W_0$ and with constant rate $\nu_x=\nu_0/\rho_x$. The kick amplitude $n$ (in units of $\Delta W_0$) follows a
binomial distribution
\begin{equation} \label{Eq:Binom-Prob1}
  p_n^{(N)} = \frac{N!}{n!(N-n)!} \rho_{x}^n (1-\rho_{x})^{N-n} \quad
\end{equation}
with average $\langle n \rangle= \rho_x N$ and variance ${\rm var}[n]=\rho_x(1-\rho_x) N$. For the balanced case
the average input current is zero and does not depend on the correlation while the current variance per unit
time $\sigma^2$ is determined by the variability of both correlated and uncorrelated kick trains: $\sigma^2 =
\Delta W_0^2 [\langle n \rangle ^2 + {\rm var}[n] + \langle n \rangle]/T_x $ with $T_x = \nu_x^{-1}$.

For large $N$ the correlated kicks can be seen as large amplitude events that are delivered at a much lower rate
${\nu}_x \ll \nu_u$ than the uncorrelated inputs that can be assimilated to an almost continuous background. The
effect of this background consists in renormalizing the bifurcation parameter according to $\bar a = a_0 \pm
\langle \Delta W \rangle /T_x $, the shift being positive (resp. negative) for $x=e$ (resp. $x=i$). The
influence of the correlated kicks is embodied in the variance that for large $N$ (at the leading order) reads as
$\sigma^2 \simeq \langle \Delta W \rangle ^2/T_x $. Thus in the high input regime the statistical properties of
the response are determined once the average amplitude of the kick $\langle \Delta W \rangle$ and $T_x$ are
known.

In the following we characterize the coherence of the neuronal response in dependence on noise strength
$\sigma^2$ and correlation $\rho_x$ (or equivalently $T_x$)~\footnote{First the correlation $\rho_x$ is chosen
and then $\sigma^2$ is fixed (independently of $\rho_x$) by selecting the appropriate number of synapses $N$.}
changing the latter from full inhibitory correlation $\rho_i=1$ to full excitatory correlation $\rho_e=1$
including the completely uncorrelated case $\rho_e=\rho_i=0$. As indication of CR we employ the occurrence of a
minimum in the coefficient of variation $R$ at intermediate noise levels. This quantity denotes the standard
deviation of the distribution of the inter-spike time intervals (ISIs) normalized by its mean $\bar T_{ISI}$ and
attains the value $0$ for a perfectly regular signal and the value $1$ for a Poissonian process.
\begin{figure}
\includegraphics*[clip,width=80mm]{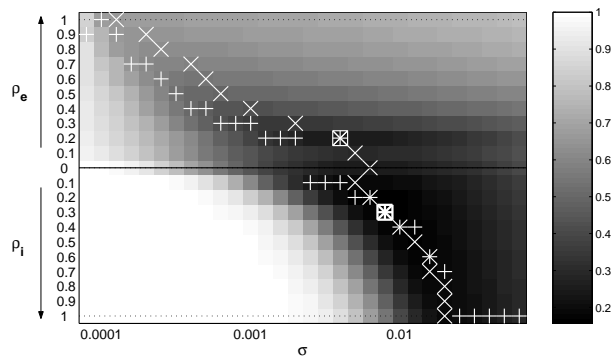}
\caption{\abb\label{fig:FHNP-CV-Im} Coefficient of variation $R$ as a function of noise strength $\sigma^2$ for
both excitatory ($\rho_e$) and inhibitory ($\rho_i$) correlation. For the sake of brevity we project the axis of
the two-dimensional plane ($\rho_e$,$\rho_i$) on one single axis. In each column a white $+$ marks the minimum
for fixed variance, while a white X in each row refers to the minimum for fixed correlation (i.e., the CRs).
Horizontal lines mark the cases with no correlation, $\rho_e=1$ and $\rho_i=1$ exhibiting CR, eCR and iCR,
respectively. Finally, the thin and thick white squares indicate the absolute minima obtained for excitatory and
inhibitory correlation, i.e., the eDCR and iDCR, respectively.  Data refer to $\nu_0=0.3$.}
\end{figure}

%
%
For fixed $\rho_x$ the coefficient of variation $R$ exhibits a minimum for intermediate values of $\sigma^2$ in
the whole range of excitatory and inhibitory correlations (cf. Fig.~\ref{fig:FHNP-CV-Im}). Depending on the type
of correlation we denote this effect as excitatory or inhibitory coherence resonance (eCR or iCR,
respectively)~\footnote{The observed CR effects can not simply be due to the dependence of the output rate on
the level of noise since for any fixed degree of correlation the rate increases strictly monotonously with the
noise strength.}. Furthermore, by ordering the correlations from full inhibitory to full excitatory for almost
all noise strengths $\sigma^2$ a minimum of $R$ can be observed at intermediate $\rho_x$. Finally, in each
half-plane $(\sigma^2,\rho_x)$ (with $x=e$ and $x=i$, resp.) an overall minimum for $R$ can be identified
corresponding to excitatory and inhibitory Double Coherence Resonance (eDCR and iDCR, respectively). The iDCR is
also the absolute minimum.

\begin{figure}
\includegraphics[clip,width=72mm,height=42mm]{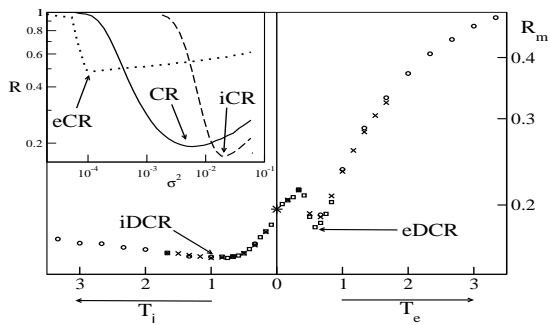}
\caption{\abb\label{fig:FHNP-CV} Minimal values of the coefficient of variation $R_m$, obtained for fixed
correlation, versus the period $T_{\rho}$. Data are ordered from full inhibitory correlation to full excitatory
correlation.  Symbols refer to different $\nu_0$-values: 0.3 (circles), 0.6 (crosses) and 1.2 (squares). The
asterisk denotes the uncorrelated case and the arrows mark the positions of the double coherence resonances.
Inset: $R$ versus $\sigma^2$ for different correlations ($\nu_0=0.3$): no correlation (solid line), $\rho_e=1$
(dotted line) and $\rho_i=1$ (dashed line). Arrows mark the position of the respective coherence resonances.}
\end{figure}

These effects can be better appreciated by considering the minimal value $R_m$ for each correlation as a
function of $T_x$ (cf. Fig.~\ref{fig:FHNP-CV}). The curves for three different $\nu_0$ almost coincide
indicating that once $T_x$ is fixed the minima $R_m$ occur at similar variances $\sigma^2_m$. Absolute minima
are observed at finite $\rho_x$ and $\sigma^2$ both for excitatory (eDCR at $\bar T_e \equiv {\bar \rho}_e /
\nu_0 \simeq 0.65$) and inhibitory correlations (iDCR at $\bar T_i \equiv {\bar \rho}_i / \nu_0 \simeq 0.9$).
This behavior of $R_m$ is accompanied by a monotonous increase of the corresponding variances $\sigma^2_m$ when
going from $\rho_e=1$ to $\rho_i=1$ (cf. Fig.~\ref{fig:FHNP-CV-Im}). As we show in the following, not only $R_m$
and $\sigma^2_m$ depend on both type and strength of correlation, but also the underlying mechanisms are
completely different from one another and in particular from the one responsible for CR without
correlation~\cite{Pikovsky97}.

\noindent \textbf{Excitatory Coherence Resonance} We begin with the origin of the \emph{eCR} at full excitatory
correlation $\rho_e=1$. For this case the ISI-distribution of the output can be described, for any $\sigma^2$,
in terms of a Poissonian process with average period $\bar T_{ISI}$ and refractory time $T_{ref}$. The
expression of $R$ is simply given by $R_{P}=1-T_{ref}/\bar T_{ISI}$, however, the course of $R$ reveals a z-like
shape with a clear minimum (the eCR, cf. inset of Fig.~\ref{fig:FHNP-CV}). At low variances firing resembles a
noise activated process and therefore $\bar T_{ISI} \gg T_{ref}$ and $R=R_P \simeq 1$. For increasing $\sigma^2$
the average kick amplitude $\langle \Delta W \rangle$ gets larger until one single EPSP may be enough to trigger
a spike, thus leading to more frequent firings and to an abrupt decrease of $R$. This quantity reaches its
minimum when a $1:1$ synchronization between the EPSPs and the output spikes sets in since now each kick is
sufficient to lead the system above the firing threshold from any state of the system (except during the
refractory period). Accordingly, for $\sigma^2 \ge \sigma^2_m$ the ISI-distribution exhibits a tail with slope
$\nu_e$ and $R= R_{sync} \equiv 1-T_{ref}/T_e$. Further increasing $\langle \Delta W \rangle$ forces the system
to fire even during the refractory period, thus leading to a reduction of $T_{ref}$ which explains the final
growth of $R$ for $\sigma^2 > \sigma^2_m$.

\begin{figure}
\centerline{\includegraphics[clip,width=80mm,height=45mm]{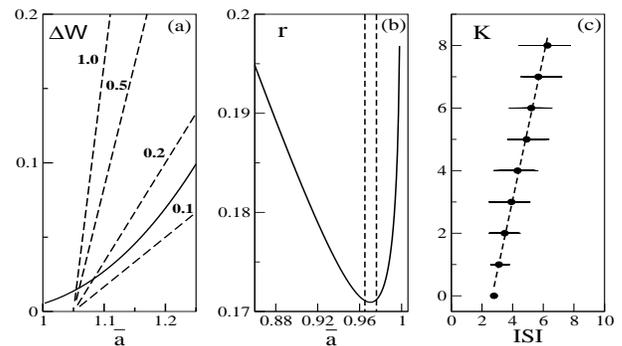}} \caption{\abb\label{fig:Mech} Excitatory
correlation: (a) Minimal amplitude $(\Delta W)_c$ (solid line) and average kick amplitude $\langle \Delta W
\rangle$ (dashed lines) versus the renormalized bifurcation parameter $\bar a$. The number in proximity of the
dashed lines indicate the corresponding values of $\rho_e$. Data refer to $\nu_0=0.3$, in this case ${\bar
\rho}_e \sim 0.2$. Inhibitory correlation: (b) Ratio $r(\bar a)$ for the deterministic FHN model, the values of
$R_m$ for $\rho_i > \bar \rho_i$ are associated with $\bar a$-values located within the two dashed lines for al
the examined $\nu_0$.(c) Number of inhibitory kicks $K$ received within a certain ISI as a function of its
duration. Filled circles mark the average for each $K$. The dashed line indicates a linear fit, the inverse of
its slope gives the delay per kick $\delta$. Data have been obtained from 6000 ISIs for $\nu_0=0.3$,
$\rho_i=0.6$, $\sigma^2=0.06$.}
\end{figure}

\noindent \textbf{Excitatory Double Coherence Resonance} This mechanism for eCR remains valid also for
decreasing correlation $\rho_e$ until $R_m$ reaches its local minimum at $\bar \rho_e$ (corresponding to the
\emph{eDCR}, cf. Fig.~\ref{fig:FHNP-CV}). Beyond the eDCR, i.e., for $\rho_e < \bar \rho_e$, the $1:1$
synchronization regime is no more reached (cf.~\ref{fig:Mech}a). For $\rho_e < \bar \rho_e$ the average kick
amplitude $\langle \Delta W \rangle$ is always smaller than the minimal amplitude $(\Delta W)_c$ needed to
elicit a spike (starting from the fixed point for the deterministic FHN, i.e., for Eq. \ref{Eq:FHN-Model-Pik}
with $I(t)\equiv0$). For $\rho_e \ge \bar \rho_e$ and sufficiently high $\sigma^2$ (or analogously, $\bar a$)
the average kick always overcomes the threshold $(\Delta W)_c$ thus confirming that the regime of $1:1$
synchronization is always reached. Since the minimum of $R$ is associated with the onset of $1:1$
synchronization (implying $R_m = R_{sync}$) $R_m$ decreases for $1 > \rho_e > {\bar \rho}_e$. For all
correlations $[\bar \rho_e; 1]$ this occurs roughly for the same amplitude $\langle \Delta W \rangle \simeq
\Delta W_c$ (cf. Fig.~\ref{fig:Mech}a), thus $T_{ref}$ is not significantly altered and the decrease of $R_m$ is
due to the reduction of $T_e$ with decreasing correlation. Finally, beyond the \emph{eDCR}, i.e., for $\rho_e <
{\bar \rho}_e$, the system is no more strictly forced by the driving kick train with very high frequency, the
period of firing decreases (i.e., $\bar T_{ISI} > T_e$) and thus $R_m$ increases.

\noindent \textbf{Inhibitory Coherence Resonance} Turning our attention to correlated inhibitory inputs we start
with examining the origin of the \emph{iCR} at full inhibitory correlation $\rho_i=1$. Once more at low
variances the emission of spikes is due to an activation process with $R \sim 1$. For increasing $\sigma^2$ the
renormalization of the bifurcation parameter, now due to the uncorrelated EPSPs, drives the system towards the
repetitive firing regime eventually crossing the bifurcation at $\bar a=1$ and this leads to a fast decrease of
$R$. On the other hand, the correlated IPSPs inhibit tonic firing, but despite the increase of their amplitude
with $\sigma^2$ their action gets less effective. This is because the fraction of time $r(\bar a)$ during which
the neuron is sensitive to the arrival of a kick before a spike emission~\footnote{This quantity can be
evaluated for the deterministic FHN as the ratio $r(a)=1-T_{ref}(a)/T_f(a)$, where $T_f(a)=1/\nu_f(a)$ is the
period of tonic firing for Eq.~(\ref{Eq:FHN-Model-Pik}) with $I\equiv 0$ and $T_{ref}(a)$ the refractory period
estimated as the time needed to recover after a spike emission from $(V,W)=(1,2/3)$ to the fixed point.}
decreases for increasing $\sigma^2$ (i.e., for smaller $\bar a$). Moreover, $r(\bar a)$ exhibits a minimum at
$\bar a \sim 0.97$. Noticeably, $R_m$ is associated with corresponding $\bar a$-values for all $\rho_i > \bar
\rho_i$ (cf. Fig. \ref{fig:Mech}b). Thus the minimum is due to the uncorrelated kick trains that renormalize the
bifurcation parameter. The final increase of $R$ at large $\sigma^2 > \sigma^2_m$ is reflected by a
"quantization" in the ISI-distribution. The single ISI is proportional to the number of inhibitory kicks
received within its duration (cf. Fig.~\ref{fig:Mech}c) and for high variances the {\it delay per kick}
$\delta$, i.e., the average retardation (with respect to the natural firing period) of the next neuronal firing
induced by a single inhibitory kick, increases and consequently the ISI-distribution broadens finally revealing
a multimodal structure.

\noindent \textbf{Inhibitory Double Coherence Resonance} This mechanism for iCR is still valid for decreasing
correlations $\rho_i$ until $R_m$ reaches its absolute minimum at $\bar \rho_i$ (corresponding to the
\emph{iDCR}, cf. Fig.~\ref{fig:FHNP-CV}). The decrease of $R_m$ with $\rho_i$ is due to the fact that $R_m$ is
associated with an almost constant $\bar a$-value within the interval $[\bar \rho_i;1]$, and that for fixed
$\bar a$ the amplitude of the inhibitory kick $\langle \Delta W \rangle \propto \rho_i$. Consequently the
average delay per kick $\delta$ decreases with the correlation until the quantization in the ISI-distribution
disappears at the iDCR at $\rho_i = \bar \rho_i$. For smaller $\rho_i \ll \bar \rho_i $ the frequency of the
inhibitory kicks becomes more important than their amplitudes. In particular, lowering the correlation towards
the uncorrelated case leads to an increasing rate $\nu_i$ of inhibitory disturbances that renders the neuronal
spiking more irregular, which in turn is reflected by the increase of $R_m$.
%
%
In summary, we described novel mechanisms for CR that are not, as in the uncorrelated case~\cite{Pikovsky97},
related to the different nature of the stochastic processes underlying neuronal firing at low and high noise but
that rely on the discrete nature of the correlated inputs. Furthermore, we reported the existence of DCRs both
for excitatory and inhibitory correlations. These DCRs reflect the change from the classical CR
~\cite{Pikovsky97} to amplitude-dominated mechanisms responsible for the CR. While the eDCR is related to a
complete synchronization with the input, the iDCR is due to a quantization of the neuronal output.

Our results indicate that the coherence in the response of an excitable system driven by fluctuations can be
modulated by controlling independently the level of correlation and the noise variance. This could be of high
relevance for neuronal coding since there are indications that correlated activities (as indeed measured, e.g.,
among cortical neurons~\cite{Zohary94}) can influence the coding ability of neuronal populations. More recently
neuronal input correlation has been linked to changes in attention suggesting its major impact on the
information flow in the brain~\cite{Salinas01}. Remarkably, in simulations with correlated inhibitory input it
has been shown that attention (modeled as an increase of synchrony in interneuron networks) can lead to a
decrease of the coefficient of variation of single output spike trains~\cite{Tiesinga04d}. Furthermore, we have
verified that the FHN with conductance-based inputs exhibits the same coherence effects as reported here for the
current-driven model. This analysis together with further extensions to more physiological setups will be the
subject of future studies. Finally, since most of the reported results are not specifically related to the
considered model we expect that analogous effects can be found also for excitable systems in other fields of
research.

We acknowledge S. Lepri and A. Politi for useful discussions and the European community for supporting TK via
the Marie Curie IEF project No 011434.


\begin{thebibliography}{99}
%
\bibitem{Gammaitoni98} L. Gammaitoni {\it et al.}, Rev Mod Phys \textbf{70} (1998) 223.

\bibitem{Gang93} Hu Gang {\it et al.}, Phys Rev Lett \textbf{71} (1993) 807.

\bibitem{Pikovsky97} A.S. Pikovsky and J. Kurths, Phys Rev Lett \textbf{78} (1997) 775.

\bibitem{Lindner04} B. Lindner {\it et al.}, Phys Rep \textbf{392} (2004) 321.

\bibitem{Manjarrez02} E. Manjarrez {\it et al.}, Neurosci Lett \textbf{326} (2002) 93.

\bibitem{FitzHugh61} R. FitzHugh, Biophys J \textbf{1} (1961) 445.

\bibitem{Buldu01} J.M. Buld{\'u} {\it et al.}, Phys Rev E \textbf{64} (2001) 051109.

\bibitem{Brugioni05} S. Brugioni {\it et al.}, Phys Rev E \textbf{71} (2005) 062101.

\bibitem{Beato05} V. Beato {\it et al.}, \pre \textbf{71} (2005) 035204(R).

\bibitem{Casado97} J.M. Casado, Phys Lett A  \textbf{235} (1997) 489.

\bibitem{Shadlen98} M.N. Shadlen and W.T. Newsome, J Neurosci \textbf{18} (1998) 3870.

\bibitem{Salinas00} E. Salinas and T.J. Sejnowski, J Neurosci {\bf 20} (2000) 6193.

\bibitem{Rudolph01} M. Rudolph and A. Destexhe, Phys Rev Lett \textbf{86} (2001) 3662.

\bibitem{Feng01} J. Feng and P. Zhang, Phys Rev E, \textbf{63} (2001) 051902.

\bibitem{Moreno02} R. Moreno {\it et al.}, Phys Rev Lett \textbf{89} (2002) 288101.

\bibitem{Rappel94} W.-J. Rappel and S.H. Strogatz, Phys Rev E \textbf{50} (1994) 3249.

\bibitem{Pradines99} J.R. Pradines, G.V. Osipov, and J.J. Collins, Phys Rev E \textbf{60} (1999) 6407.

\bibitem{Zohary94} E. Zohary, M.N. Shadlen, and W.T. Newsome, Nature \textbf{370} (1994) 140.

\bibitem{Salinas01} E. Salinas and T.J. Sejnowski, Nature Rev Neurosci \textbf{2} (2001) 539.

\bibitem{Tiesinga04d} P.H.E. Tiesinga {\it et al.}, J. Physiol. \textbf{98} (2004) 296.
%
%
%
%
%
%
%
%
%
%
%
%
%
%
%
%
%
%
%
%
\end{thebibliography}
\end{document}